\documentstyle[12pt]{article}
\begin{document}
\normalsize
\def\vec#1{{\mbox{\boldmath $#1$}}}  
\def\refname{}   
\def\bi#1 {\vspace{-0.5\baselineskip} \bibitem{#1}}
\def\D {Delbr\"uck {}}
\def\sub#1{{\mbox{\scriptsize #1}}}

\def\beq {\begin{equation}}
\def\eeq {\end{equation}}
\def\beqn {\begin{eqnarray}}
\def\eeqn {\end{eqnarray}}
\def\ee{(\vec e\vec e')}
\def\ss{(\vec s\vec s')}

\def\simge{\hspace*{0.2em}\raisebox{0.5ex}{$>$}
     \hspace{-0.8em}\raisebox{-0.3em}{$\sim$}\hspace*{0.2em}}  
\def\simle{\hspace*{0.2em}\raisebox{0.5ex}{$<$}
     \hspace{-0.8em}\raisebox{-0.3em}{$\sim$}\hspace*{0.2em}}  

\normalsize
\begin{center}
{\Large \bf Budker Institute of Nuclear Physics}
\end{center}
\begin{flushright}
BINP 94-30\\
June 1994
\end{flushright}
\vspace{1.0cm}

\begin{center}{\Large \bf Relativistic Oscillator Model and \D scattering}\\
\vspace{1.0cm}

 A.I. L'vov\\ \it
 Lebedev Physical Institute,\\ \it
 Leninsky Prospect 53, 117924 Moscow, Russia
 \\ ~ \\
 A.I. Milstein\\ \it
 Budker Institute of Nuclear Physics,\\ \it
 630090 Novosibirsk, Russia
 \\ ~ \\

\end{center}

\begin{abstract}
Elastic scattering of photons in a Lorentz-scalar potential
via virtual spin-zero particle-antiparticle pairs
(``\D scattering") is considered. An analytic expression
for the \D amplitude is found exactly in case of an oscillator
potential. General properties of the amplitude and its asymptotics
are discussed.
\end{abstract}

\newpage
\bigskip {\bf 1.~}
Elastic scattering of photons in the Coulomb field of nuclei via
virtual electron-positron pairs (\D scattering \cite{LM})
attracted considerable interest for a long time. That is motivated by two
reasons: (i) The \D scattering is one of a few nonlinear
quantum-electrodynamic processes which can be precisely tested by experiment
\cite{PM,J,D15,D13}. (ii) In order to extract an information on
nuclear structure from differential cross sections of photon scattering
on nuclei, a precise knowledge of the \D amplitude may be required
because of its interference with the nuclear amplitude. In some cases, the \D
scattering considerably modifies the differential cross section
of $\gamma$A-scattering \cite{D19,D29,D30,D37}.

Theoretical investigations of the \D scattering have a long history and
many papers are devoted to this subject. Now the \D amplitude is studied
in detail in some approximations: (1) In the
lowest-order Born approximation with respect to the parameter $Z\alpha$
(here $Z|e|$ is the charge of the nucleus, $\alpha = e^2 \simeq 1/137 $ is the
fine-structure constant), results were obtained for an arbitrary
momentum transfer $q$; these results are surveyed in detail in
\cite{PM}. (2) For the case of high energies
($\omega \gg m_e$, $m_e$ being the electron mass) and small scattering angles
($q \ll \omega$), the \D amplitude was obtained in
\cite{CW1,CW2,CW3,MS1,MS2} to all orders in $Z\alpha$.
It was found that the Coulomb corrections at $Z\alpha \sim 1$
drastically change the amplitude as compared to the Born approximation.
(3) At high energies and momentum transfers ($\omega,q \gg m_e$),
the amplitude was also found to all orders in $Z\alpha$
\cite{MSh,MRS1,MRS2}.  In this case the \D amplitude has a scaling behavior
\cite{LB4} and becomes inversely proportional to the photon energy.
It has been shown that the Coulomb corrections essentially decrease the
amplitude at high momentum transfer as well.
Recently some general exact expressions for the \D amplitude were derived
in \cite{G}, although numerical results were not presented.
Some numerical results for the \D amplitude obtained in the lowest-order
Born approximation, in the "high-energy and small-angle" approximation,
and in the "high energy and momentum transfer" approximation
can be found in \cite{MRS3}.

In the present paper we consider the \D scattering in the scalar QED
in the field of an oscillator Lorentz-scalar potential, or, in other words,
for a relativistic oscillator. The relativistic oscillator model,
being a nice theoretical laboratory, has also a few realistic applications.
For instance, it can be successfully used \cite{LvMi}
to describe interactions of collective modes of a nucleus
in the region of giant resonances with the electromagnetic field.
It takes the advantage of being automatically consistent
with microcausality, analyticity, and dispersion relations.
The \D amplitude is a quantum correction to a classical part of the
photon scattering by the oscillator and it is interesting to evaluate
this correction explicitly to understand its significance for the physics
of photon-nucleus scattering. Besides, the \D amplitude for the relativistic
oscillator provides an example of exact calculation
of this quantity in external potentials.

\bigskip {\bf 2.~}
To be specific, we consider spin-zero particles described by
the Klein-Gordon equation with an oscillator Lorentz-scalar potential.
We start with the Lagrangian for the charged quantum field $\hat\phi$
in the external electromagnetic potential $A_\mu$~:
\beq\label{eq:Lag}
 L[\hat\phi,A](x) = \left| \partial_\mu \hat\phi(x) +
    ie A_\mu(x) \hat\phi(x)\right|^2 - U(r)|\hat\phi(x)|^2,
    \quad U(r)=\mu^2 + \gamma^4  r^2.
\eeq
In the Furry representation, the Feynman rules corresponding to
this Lagrangian are found from the usual Feynman rules for
spin-zero particles by replacing plane waves with the normalized solutions
$\langle\vec r|n\rangle = \phi_n(\vec r)$
of the Klein-Gordon equation in the potential $U(r)$~:
\beq\label{eq:KG}
  p^2 \phi_n(\vec r) + \gamma^4 r^2 \phi_n(\vec r)
    = (E_n^2 - \mu^2) \phi_n (\vec r), \quad \vec p=-i\vec\nabla
\eeq
(for brevity, we do not indicate here quantum numbers related to
the angular momentum).
The equation (\ref{eq:KG}) coincides with that for a nonrelativistic
oscillator and has the spectrum $E=\pm E_n$, where
\beq\label{eq:En}
  E_n=\sqrt{\mu^2 + (2n+3) \gamma^2}, \quad n\ge 0.
\eeq
The parameter $\gamma^4$ determines a slope of the potential $U(r)$
in (\ref{eq:Lag}) and must be positive to result in a stable vacuum
and bound states. It also determines a range of the ground state,
$\phi_0(r) = \gamma^{3/2}\pi^{-3/4} \exp(-\frac12 \gamma^2 r^2)$.
The parameter $\mu^2$ may be negative.
However, we require $\mu^2 > -3\gamma^2$ to have a positive energy gap
$2E_0$ between levels of positive and negative energies and hence a
stable vacuum.
In the nonrelativistic limit, $\gamma \ll \mu$,
the parameter $\mu$ becomes the mass $m$
of the particle and the oscillator parameter $\gamma$ determines
the oscillator frequency $\omega_0=E_{n+1} - E_n=$const,
\beq
 \mu \to m, \quad  \gamma^2 \to m\omega_0 .
\eeq

The \D amplitude for the spin-zero particle
is described by two Feynman diagrams shown in Fig.~1.
In this figure, double lines represent the Green function for the
Klein-Gordon equation in the external potential $U(r)$. The first diagram,
Fig.~1a, describes the so-called seagull contribution to the photon
scattering amplitude which is equal to
\beq\label{eq:SDdef}
 S_D = -2i\alpha\ee \int_{-\infty}^\infty \frac{d\epsilon}{2\pi}\int
    d^3r \, G(\vec r,\vec r | \epsilon)\,\exp(i\vec q\vec r) =
  -\alpha\ee \sum_n \frac1{E_n} \langle n|\exp(i\vec q\vec r)|n\rangle.
\eeq
Here $\vec e$ ($\vec k$) and $\vec e'$ ($\vec k'$) are
polarizations (momenta) of the incoming and outgoing photons,
respectively, and $\vec q= \vec k-\vec k'$~;
we use the radiative gauge, $\vec e\vec k=\vec e'\vec k'=0$.
$G(\vec r,\vec r'|\epsilon)$ is the Green function in the potential
$U(r)$:
\beq\label{eq:GFdef}
  G(\vec r,\vec r'|\epsilon) = \sum_n \frac
 {\langle\vec r|n\rangle \langle n|\vec r'\rangle}
  {\epsilon^2-E_n^2+i0}
    = -i \int_0^\infty ds \,\langle\vec r|
      \exp\left[is(\epsilon^2 -p^2-U(r)+i0)\right]|\vec r'\rangle,
\eeq
$s$ being a proper time. The amplitude we use is normalized as to give,
being squared, exactly the differential cross section of photon scattering.

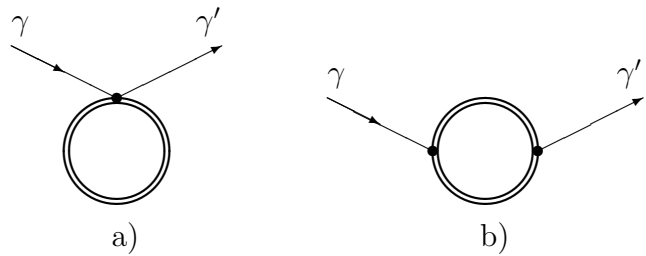
\begin{figure}[ht]
\begin{center}
\unitlength 0.7mm
\begin{picture}(140,40)(0,15)

\thicklines
\put(30,30){\circle{19}}
\put(30,30){\circle{20}}
\put(30,40){\circle*{2}}
\thinlines
\put(10,50){\line(2,-1){20}}
\put(10,50){\vector(2,-1){10}}
\put(10,53){\makebox{$\gamma$}}
\put(30,40){\vector(2, 1){20}}
\put(45,53){\makebox{$\gamma'$}}
\put(29,12){\makebox{a)}}

\thicklines
\put(100,30){\circle{19}}
\put(100,30){\circle{20}}
\put( 90,30){\circle*{2}}
\put(110,30){\circle*{2}}
\thinlines
\put( 70,40){\line(2,-1){20}}
\put( 70,40){\vector(2,-1){10}}
\put( 70,43){\makebox{$\gamma$}}
\put(110,30){\vector(2,1){20}}
\put(125,43){\makebox{$\gamma'$}}
\put( 99,12){\makebox{b)}}

\end{picture}
\end{center}
\caption{Diagrams of the \D scattering: (a) Seagull amplitude $S_D$.
  (b) Resonance amplitude $R_D$.}
\end{figure}

Using a known expression for the time-dependent Green function
of a nonrelativistic particle in an oscillator potential
(see e.g. \cite{FH}), we obtain an explicit form for the Green function
(\ref{eq:GFdef}):
\beqn\label{eq:GF}
  && G(\vec r,\vec r'|\epsilon) = -i\gamma\int_0^\infty
  \frac{ds}{(2\pi i\sin 2s)^{3/2}}\times \nonumber\\ && \qquad
  \qquad \exp\left\{ \frac{is}{\gamma^2}(\epsilon^2-\mu^2+i0)
  +\frac{i\gamma^2}{2\sin 2s}
  \left[(r^2+{r'}^2)\cos 2s - 2\vec r\vec r'\right] \right\}.
\eeqn
Substituting (\ref{eq:GF}) to (\ref{eq:SDdef}) and taking elementary
integrals with respect to $\epsilon$ and $\vec r$, we find:
\beq\label{eq:SD}
 S_D = -\frac{i\alpha e^{i\pi/4}}{8\gamma\sqrt\pi}\ee\int_0^\infty
 \frac{ds}{\sqrt s\sin^3 s}\,
 \exp\left[-is\frac{\mu^2-i0}{\gamma^2}
 + \frac{iq^2}{4\gamma^2}\cot s \right].
\eeq
In case of the oscillator potential $U(r)$, the Klein-Gordon operator
has only a discrete spectrum (\ref{eq:En})
and therefore the Green function has only simple poles
at real $\epsilon=\pm (E_n-i0)$ displaced by $i0$
in accordance with the Feynman rules.
Using this analytic property, we can deform the integration path
over $\epsilon$ in (\ref{eq:SDdef}),
$\epsilon\to i\epsilon$, to make it finally coincident
with the imaginary axis. As a result, we get a possibility to
rotate the contour of integration over $s$ in
(\ref{eq:GFdef})--(\ref{eq:SD}),
$s\to -is$, to make all the integrals well-convergent.

Note that an ultraviolet divergence at $s\to 0$ is not present in
the amplitude $S_D$ provided $q \neq 0$, so that a regularisation is
not necessary. The same is valid for the second diagram, Fig.~1b,
which gives the so-called resonance contribution:
\beqn\label{eq:RDdef}
 R_D &=& -4i\alpha\int_{-\infty}^\infty \frac{d\epsilon}{2\pi}
  \int d^3r \, d^3r'
  \Big[i\vec e \vec\nabla G(\vec r,\vec r'|\epsilon)\Big]
  \times\nonumber \\ &&\qquad\qquad
  \Big[i\vec e'\vec\nabla'G(\vec r',\vec r|\epsilon +\omega)\Big]
  \,\exp(i\vec k\vec r - i\vec k'\vec r')  \\
 \label{eq:RDpoles} &=&
  \alpha\sum_{n,n'} \frac{
 \langle n |(\vec e'\vec p)\exp(-i\vec k'\vec r)|n'\rangle
 \langle n'|(\vec e \vec p)\exp( i\vec k \vec r)|n \rangle }
 {E_n E_{n'} (E_n + E_{n'} -\omega -i0)} +(\omega\to -\omega).
\eeqn
It has a series of poles at $\omega=\pm(E_n+E_{n'})$ which are related
with the pair photoproduction from vacuum when the pair components
are captured to the levels $n$, $n'$ of the discrete spectrum.
The pole at $\omega=2E_0$ is
nevertheless absent because matrix elements in (\ref{eq:RDpoles})
vanish for the $s$-wave states $n=n'=0$.
Using (\ref{eq:GF}), we easily calculate the integrals (\ref{eq:RDdef})
with respect to the variables $\epsilon$,
$\vec r$ and $\vec r'$ (here the substitutions $\vec r=\vec r_1+\vec r_2$
and $\vec r'=\vec r_1-\vec r_2$ are helpful).
It is also convenient to replace
$s_1 = s(1+x)/2$ and $s_2=s(1-x)/2$, where $s_1$ and $s_2$
are proper times in the integral representation of the
Green functions in (\ref{eq:RDdef}). Then we get
\beqn\label{eq:RD}
  && R_D = -\frac{\alpha e^{i\pi /4}}{8\gamma\sqrt\pi}
  \int_0^\infty ds \int_0^1 dx\,
  \frac{\sqrt s}{\sin^4 s}
  \left[ \,\frac{\sin^2 (sx)}{2\gamma^2\sin s}
  (\vec e\vec k')(\vec e'\vec k)
  -i\cos(sx)\ee \right] e^{iF},
\eeqn
where
\beq\label{eq:F}
 F = -s\frac{\mu^2-i0}{\gamma^2}+(1-x^2)\frac{\omega^2 s}{4\gamma^2}
   + \frac{q^2-2\omega^2}{4\gamma^2 \sin s} \cos (sx)
   +\frac{\omega^2}{2\gamma^2}\cot s ~.
\eeq
Integrating the term in (\ref{eq:RD}) proportional to $\ee$
by parts with respect to the variable $x$, we find
that the term outside the integral cancels the contribution $S_D$ of
the first diagram, so that the \D scattering amplitude $T_D=S_D+R_D$
finally reads:
\beq\label{eq:TD}
  T_D= -\frac{\alpha e^{i\pi/4}\omega^2}{16\gamma^3\sqrt\pi}
  \int_0^{\infty}ds\int_0^1 dx\,
  \frac{\sqrt s \sin(sx)}{\sin^5 s}
  \left[x\sin s\,\ee  - \sin(sx)\ss \right] e^{iF},
\eeq
where $\ss\equiv (\vec{\hat k}\times\vec e)(\vec{\hat k}'\times\vec e')$
describes a magnetic response.
We see that a contribution like $S_D$ which depends on $q$ and is
independent of $\omega$ vanishes in the total amplitude $T_D$, as it
must be according to general consequences of gauge invariance and related
low-energy theorems. Eq.~(\ref{eq:TD}) is our main result.

We may obtain another form of Eq.~(13) by deforming
the integration path over $\epsilon$ in (\ref{eq:RDdef})
to transform the integral to
$\int_{-\frac12\omega-i\infty}^{-\frac12\omega+i\infty} d\epsilon$.
Such a deformation is always possible at low energies $\omega$
because the chains of singularities of two Green functions in (\ref{eq:RDdef})
lying below and above the real axis, i.e. at
$\epsilon=E_n-\omega-i0$ and at $\epsilon=-E_n+i0$, respectively,
do not pinch the integration path provided $\omega < 2E_0$.
Then we can rotate $s_1\to -is_1$, $s_2\to -is_2$ and get
well-convergent integrals. Respectively, we may rotate
the integration path in (\ref{eq:TD}), $s\to-is$, and arrive at a real
integral which is well suitable for calculating $T_D$ at low energies.
However, such an integral turns out to be divergent at
$\omega>E_0+E_1$, i.e. above the nearest pole of the amplitude $T_D$,
see Eq.~(\ref{eq:RDpoles}).

\bigskip {\bf 3.~}
Eq.~(\ref{eq:TD}) can be further simplified in case of a low
oscillator frequency and momentum transfer,
$\gamma^2\ll \mu^2$ and $q\ll \mu$.
Then the contribution of the region $s\simge 1$ to the integral
in (\ref{eq:TD}) is exponentially suppressed, as is seen after the
rotation of the contour, whereas the
contribution from the region $s\ll 1$ reads
\beqn\label{eq:TDas}
  && T_D\simeq -\frac{\alpha e^{i\pi/4}\omega^2}{16\gamma^3\sqrt\pi}
  [\ee-\ss] \int_0^\infty \frac{ds}{s^{5/2}} \int_0^1 x^2 dx\times
  \nonumber\\ && \qquad\qquad
  \exp\left[ -is\frac{\mu^2-i0}{\gamma^2} +i\frac{q^2}{4\gamma^2 s} -
   i\frac{(1-x^2)^2}{48\gamma^2} \omega^2s^3\right]
\eeqn
and is saturated by
\beq
    s\sim s_\sub{eff} =\cases{
    q^2\gamma^{-2}, & if ~ $\mu q\simle\gamma^2$ \cr
    q\mu^{-1}, & if ~ $\mu q\simge\gamma^2$.}
\eeq
Respectively, at ``low" energies $\omega^2\ll\gamma^2 s_\sub{eff}^{-3}$
which may nevertheless be high in comparison with $\mu$,
the integral (\ref{eq:TDas}) reduces to the modified Bessel function
of the third kind $K_{3/2}$ and hence is an elementary function:
\beq\label{eq:TDlow}
 T_D\simeq \frac{\alpha\omega^2}{12 q^3}
  \left(1+\frac{\mu q}{\gamma^2}\right)
  \exp\left(-\frac{\mu q}{\gamma^2}\right) [\ee-\ss].
\eeq
When $q\to 0$, the \D amplitude becomes proportional to an effective
volume of the vacuum, $V\sim q^{-3}$, probed by the photons.
At high energies $\omega^2\gg \gamma^2 s_\sub{eff}^{-3}$,
the integral (\ref{eq:TDas}) is saturated by $x$ close to 1 and the answer
is given by the modified Bessel function $K_3$:
\beq\label{eq:TDhigh}
 T_D\simeq \pm i\omega\sqrt3 \frac{\alpha\mu^3}{\gamma^2 q^3}
  K_3\!\left(\frac{\mu q}{\gamma^2}\right) [\ee-\ss],
\eeq
where the sign of plus or minus appears for Im\,$\omega>0$ or
Im\,$\omega<0$, respectively. Since $K_3(z)\to 8z^{-3}$ when $z\to 0$,
the last expression seems to have a too strong singularity
$\sim q^{-6}\sim V^2$ when $q\to 0$.
However, when $q$ is very small, the assumption
$\omega^2\gg \gamma^2 s_\sub{eff}^{-3}$
becomes violated and the \D amplitude approaches to the regime of
Eq.~(\ref{eq:TDlow}).

The singular behavior of the \D amplitude at small $q$,
Eq.~(\ref{eq:TDlow}), can be understood in the following way.
If the external potential $U$ is equal to zero,
the photon scattering amplitude due to the diagrams shown in Fig.1
reads \cite{Schw51}:
\beq
  T=  (2\pi)^3 \delta^3(\vec k-\vec k')
   \frac{\alpha}{48\pi^2} {e'}^\mu e^ \nu (k_\mu k_\nu - k^2 g_{\mu\nu})
   \log \frac{\Lambda^2}{m^2_0},
\eeq
where $\Lambda$ is an ultraviolet cut off and $m_0$ is the mass of bare
scalar particles, and is proportional to $\omega^2[\ee-\ss]$.
That means that the vacuum has an electric and
magnetic polarizability per unit volume,
\beq
  \chi_E^\sub{vac,0}=-\chi_M^\sub{vac,0} =
        \frac{\alpha}{48\pi^2}  \log \frac{\Lambda^2}{m^2_0}.
\eeq
These universal vacuum susceptibilities are absorbed by renormalisation
of electromagnetic fields and charges and are not observable.
However, in the presence of an almost uniform scalar potential $U(r)$ which
shifts the mass of the particles $m^2_0\to m^2_\sub{eff}\equiv U(r)$,
the polarizabilities get a finite meaningful piece,
\beq\label{eq:Vac}
   \chi_E^\sub{vac}(r) = -\chi_M^\sub{vac}(r) =
        -\frac{\alpha}{48\pi^2}  \log \frac{U(r)}{m^2_0}.
\eeq
The last formula is valid when the potential $U(r)$ is constant at
distances $\sim m^{-1}_\sub{eff}$ which are characteristic
for creating those vacuum polarizabilities:
\beq
   m^{-1}_\sub{eff} \ll \Delta r \equiv U(r)
   \left[\frac{dU(r)}{dr}\right]^{-1}.
\eeq
For the potential $U(r)=\mu^2+\gamma^4 r^2$, it means
\beq
  \cases{  r\gg \gamma^{-1}, &  if ~ $\gamma^2 r \simge \mu$ \cr
           r\ll \mu^3\gamma^{-4}, & if ~ $\gamma^2 r \simle\mu$}
\eeq
In the case $\gamma \ll \mu$, both the above regions
overlap and the formula (\ref{eq:Vac}) is valid everywhere.
Then the scattering amplitude of low-energy photons is equal to
\beq\label{eq:TDalpha}
    T_D= \omega^2\alpha_E(q)\ee + \omega^2\alpha_M(q)\ss
\eeq
with
\beq\label{eq:alpha}
    \alpha_E(q) = - \alpha_M(q) = -\frac{\alpha}{48\pi^2} \int
    \exp (i\vec q\vec r) \log \frac{U(r)}{m^2_0} \,d^3r ~.
\eeq
Taking this elementary integral, we find
\beq\label{eq:EM}
    \alpha_E(q) = - \alpha_M(q) =
    C \delta^3(\vec q) + \frac{\alpha}{12q^3}(1+a)e^{-a},
    \quad a=\frac{\mu q}{\gamma^2},
\eeq
in complete accordance with (\ref{eq:TDlow}).
Here $C$ is an infinite constant due to the polarizability of the
vacuum in the whole space; it is infinite because the potential
$U(r)\to\infty$ when $r\to\infty$.

The approximation of a uniform potential (\ref{eq:Vac}) turns out to
be inapplicable at high energies when the virtual particles of the mass
$m_\sub{eff}$ produced by the photon propagate to a distance
$\sim \hbar c / \Delta E \sim \omega / m^2_\sub{eff}$
which is large in comparison with the scale $\Delta r$
of variation of the potential $U(r)$.
This just happens when $\omega^2\gg \gamma^2 s_\sub{eff}^{-3}$
and the amplitude $T_D$ becomes predominantly imaginary,
see Eq.~(\ref{eq:TDhigh}).

Keeping in (\ref{eq:TD}) terms of the next order in $s^2$, we can
find a correction to the expansion of $T_D$ in powers of
the momentum transfer. At ``low" energies it reads
\beq\label{eq:TDnext}
  T_D \simeq \mbox{Eq. (\ref{eq:TDlow})} +
   \frac{\alpha\omega^2 e^{-a}}{1440\mu\gamma^2}
   [(19+2a)\ss - (17+2a)\ee] ~.
\eeq
It determines an asymptotics of the helicity-non-flip amplitude
$T_D^{++}$ because the piece (\ref{eq:TDas})
contributes to only the helicity-flip amplitude $T_D^{+-}$, as easily
seen from the relation $\ee = \pm\ss = (1\pm\cos\theta)/2$ for the
helicity-non-flip and helicity-flip case, respectively,
$\theta$ being the scattering angle.

The singular behavior of $T_D$ at $q\to 0$ disappears when the
potential $U(r)$ has a finite range. For example, in case of a cut
oscillator potential,
\beq
  \log\frac{U(r)}{m^2_0} = -A\exp\left(-\frac{\omega^2_0 r^2}{A}\right),
\eeq
where $\displaystyle A=\log\frac{U(\infty)}{U(0)}
   = O\left(\frac{\omega_0}{m_0}\right)\ll 1$
and $\omega_0$ is the frequency of small oscillations,
the polarizabilities (\ref{eq:alpha}) are finite:
\beq
  \alpha_E(q)=-\alpha_M(q)=\frac{\alpha A^{5/2}}{48\sqrt\pi\omega^3_0}
  \exp\left(-\frac{Aq^2}{4\omega^2_0}\right).
\eeq
At energies $\omega=O(\omega_0)$ the corresponding \D amplitude
(\ref{eq:TDalpha}) is less by the factor of $O((\omega_0/m_0)^{3/2})$
than the ordinary nonrelativistic photon scattering amplitude
$\displaystyle T_0\simeq \frac{\alpha\omega^2}{m_0(\omega^2_0-\omega^2)}\ee$
by a particle bound at the ground state.

\bigskip {\bf 4.~}
In conclusion, in the present paper we calculated for the first time
the amplitude of the \D scattering in the scalar QED in case of
a scalar oscillator potential and, at low oscillator frequency,
investigated its asymptotic behavior at low and high energies
and low momentum transfer.
A close relation of the \D scattering with the vacuum polarisation
was demonstrated.

We are pleased to thank M.~Schumacher and D.~Drechsel for hospitality
at University of G\"ottingen and University of Mainz where a part of
this work was done (by A.~M. and A.~L., respectively).
This work was supported by the Russian Fund for Fundamental Research,
by the Deutsche Forschungsgemeinschaft, and by the WE-Heraeus-Stiftung.

\end{document}